\documentclass[12pt]{article}
\usepackage[top=0.5in, right=0.4in, left=0.4in, bottom=0.8in]{geometry}
\usepackage{amsmath, amsfonts, amsthm, amssymb, enumerate, color, authblk, color, bm, graphicx, multirow, comment, url, xspace, natbib, titlesec, hyperref, ulem, algpseudocode}
\usepackage{amsmath, amsfonts, amsthm, amssymb, enumerate, color, authblk, color, bm, graphicx, multirow, comment, url, xspace, natbib, titlesec, hyperref, ulem, lscape,float}
\usepackage{hhline}
\usepackage{amsmath}
\usepackage{graphicx}
\usepackage{makecell}
\usepackage{cleveref}
\usepackage{lscape}
\usepackage{pdflscape}
\usepackage{longtable}
\usepackage{rotating}
\usepackage{graphicx}
\usepackage{adjustbox}
\usepackage{ upgreek } 
\usepackage{fullpage}
\usepackage{times}
\usepackage{appendix}
\usepackage[linesnumbered,ruled,vlined]{algorithm2e}

\newcommand{\ie}{{\it i.e., }}
\newcommand{\eg}{{\it e.g., }}

\newcommand{\comments}[1]{}
\newcommand{\removed}[1]{}

\begin{document}
\title{Comparing Mixture, Box, and Wasserstein Ambiguity Sets in Distributionally Robust Asset Liability Management}




\author[1]{Alireza Ghahtarani\thanks{Corresponding author (alireza.ghahtarani@hec.ca).}}
\author[2]{Ahmed Saif \thanks{Contributing author (Ahmed.Saif@dal.ca).}}
\author[2]{Alireza Ghasemi \thanks{Contributing author (alireza.ghasemi@dal.ca).}}
\affil[1]{Department of Logistics and Operations Management, HEC Montréal}
\affil[2]{Department of Industrial Engineering, Dalhousie University}

\date{}

\maketitle

\abstract{Asset Liability Management (ALM) represents a fundamental challenge for financial institutions, particularly pension funds, which must navigate the tension between generating competitive investment returns and ensuring the solvency of long-term obligations. To address the limitations of traditional frameworks under uncertainty, this paper implements Distributionally Robust Optimization (DRO), an emergent paradigm that accounts for a broad spectrum of potential probability distributions. We propose and evaluate three distinct DRO formulations: mixture ambiguity sets with discrete scenarios, box ambiguity sets of discrete distribution functions, and Wasserstein metric ambiguity sets. Utilizing empirical data from the Canada Pension Plan (CPP), we conduct a comparative analysis of these models against traditional stochastic programming approaches. Our results demonstrate that DRO formulations, specifically those utilizing Wasserstein and box ambiguity sets, consistently outperform both mixture-based DRO and stochastic programming in terms of funding ratios and overall fund returns. These findings suggest that incorporating distributional robustness significantly enhances the resilience and performance of pension fund management strategies.}

\section{Introduction}\label{sec1paper2}

Asset-liability management (ALM) refers to the challenge of managing the assets and liabilities of an entity in a way that ensures the entity can meet its financial obligations in the future \citep{zenios1995asset}. The ALM problem typically arises in financial institutions such as banks, insurance companies, and pension funds, which have significant liabilities that must be met over a long period of time. In particular, ALM is a critical concern for pension funds that must ensure they meet specific obligations to individuals who have contributed to their funds while also generating investment returns \citep{bodie1988defined}. 

Pension funds play a crucial role in the global financial landscape as evidenced by their substantial assets which exceeded $\$60.6$ trillion by the end of 2021, accounting for $33\%$ of global assets \footnote{https://www.thinkingaheadinstitute.org/research-papers/global-pension-assets-study-2022/}. This magnitude is exemplified by the fact that in nine out of the 38 Organisation for Economic Co-operation and Development (OECD) countries, pension fund assets surpassed their respective GDPs. In the last decade (2010-2020), pension assets have grown by $5.7\%$ \footnote{https://www.statista.com/statistics/721151/average-growth-largest-pension-markets-worldwide/}, outpacing the $2.6\%$ GDP growth rate over the same period \footnote{https://www.macrotrends.net/countries/WLD/world/gdp-growth-rate} and underscoring the increasing importance of retirement savings worldwide. However, as the aging population grows, outflows from pension funds to cover benefits are also accelerating. The ratio of benefits paid from retirement savings plans to GDP varies across OECD countries, ranging from $0.5\%$ to $8\%$ \footnote{https://www.oecd.org/finance/private-pensions/globalpensionstatistics.htm}. This growth in both assets and payouts highlights the need for prudent and sustainable management of pension funds to ensure that retirees receive their benefits without putting undue stress on the funds' assets.

Pension funds are essential for ensuring retirement income security; however, they encounter challenges arising from demographic changes, low-interest rates, and increasing life expectancy. To address these challenges, reforms have been implemented, including raising the retirement age and promoting private pension plans \citep{holzmann2013global}. However, one of the major challenges for pension fund managers is the uncertainty surrounding future asset returns and liabilities. The asset returns and the value of liabilities can fluctuate due to factors like inflation, interest rates, and market conditions. To mitigate this risk, effective ALM strategies are necessary. These strategies involve monitoring and managing investment portfolios to ensure they are well-aligned with future obligations. By optimizing the ALM problem under uncertainty, pension funds can enhance long-term sustainability and avoid financial distress \citep{gulpinar2016robust}.

Among the powerful techniques for managing uncertainty in ALM problems is stochastic programming (SP), which explicitly models uncertainty through probability distributions of asset return and liability values and finds optimal asset allocation strategies of the portfolio under different scenarios. This approach allows investors to better hedge against unexpected changes in asset returns and liability values, while still maintaining a desirable level of return \citep{kouwenberg2001scenario}. SP can also be used to optimize dynamic asset allocation strategies, where the asset allocation is periodically adjusted in response to changing market conditions and liability values \citep{consigli1998dynamic, dempster1996dynamic, hibiki2006multi}. By modeling the stochastic behavior of asset returns and liability values, investors can make more informed decisions about the optimal timing and size of asset allocation adjustments, and better manage the risk of underfunding future liabilities. For more details on the application of SP in ALM problems, we refer to \citep{klaassen1997discretized, kouwenberg2001scenario, consigli2008asset, duarte2017asset, kopa2018individual, barro2022stochastic}. Despite its intuitive appeal and favorable convergence properties, SP requires large amounts of data on asset returns and liability values to construct their probability distributions, which may be difficult to obtain or may not be available. Furthermore, SP is a risk-neutral approach and thus does not provide sufficient protection against adverse scenarios.

Another popular framework for dealing with uncertainty is robust optimization (RO), which seeks to find solutions that perform optimally under worst-case scenarios, in contrast to SP that aims to optimize the expected performance \citep{ben2009robust, gabrel2014recent, ghahtarani2022robust}. In the context of ALM, RO can be used to find an asset allocation strategy that is most robust to uncertainty in asset returns and liability values. 

A few recent attempts have been made to apply RO to mitigate uncertainty in the ALM problem. 
\cite{iyengar2016robust} introduced a robust factor model to capture the true uncertainty of asset returns in pension fund management. By incorporating a factor model with stochastic parameters, they developed an ALM formulation with a constraint on the funding ratio. The funding ratio, representing the assets' value relative to the present value of liabilities, is subject to uncertainty. The proposed formulation assumes the funding ratio as an uncertain parameter and utilizes a Gaussian process for factors. \cite{platanakis2017asset} extended this approach by considering ellipsoidal uncertainty sets for factor loading, box ambiguity sets for asset returns and liabilities, and upper and lower bounds for the covariance matrix of disturbances, which enabled the problem to be reformulated as a second-order cone programming (SOCP) model. Based on the results, these robust factor models and formulations enhance the out-of-sample performance of ALM problems. 
Alongside robust factor models for ALM problems, \cite{gulpinar2013robust} proposed a robust ALM using time-varying investment opportunities. They extended the multiperiod PSP formulation of \cite{dantzig1993multi} by including liabilities and funding ratio constraints. In this formulation, cumulative rates of return of assets are treated as uncertain parameters within an ellipsoidal uncertainty set. Moreover, asset returns and interest rates are modeled by using the vector-autoregressive process to capture the dynamic nature of investments. In contrast to other robust ALM approaches, \cite{gulpinar2016robust} developed an asymmetric uncertainty set to better reflect the actual uncertainty structure. \cite{gajek2022robust} proposed a robust ALM formulation with uncertain interest rates, where the distribution function of the uncertain parameters belongs to a nonempty ambiguity set. This formulation provides an upper bound on VaR (Value at Risk) for portfolio value changes caused by violations of the interest rate model. Finally, the ALM problem with discrete recourse decision and parameter uncertainty has been addressed by \cite{ghahtarani2023double} using the $K$-adaptability approach.
However, despite being a risk-averse and distribution-free approach, RO usually results in overly conservative investment strategies, which can lead to missed opportunities for higher returns, thus negatively impacting the long-term performance of pension funds. 

Disadvantages of the application of either SP or RO in ALM problems provide motivation for the application of a relatively new approach to ALM problems called distributionally robust optimization (DRO) \citep{rahimian2019distributionally}. Like RO, DRO aims to minimize the impact of uncertain scenarios on investment decisions. However, DRO goes one step further by enabling the available information about the probability distribution of random variables, albeit limited and imperfect, to be incorporated into the decision-making process, thus leading to less conservative and more stable investment strategies \citep{lin2022distributionally}. 
Various researchers, such as \citep{jiang2023distributionally, zhang2023high, blanchet2022distributionally, zhang2023high}, have employed DRO in addressing the PSP. However, these implementations do not tackle the ALM problem, as they have not accounted for liabilities and uncertainties of liabilities within their models. Unlike SP and RO, which have been applied to the ALM problem, DRO is yet to be leveraged in this context. One reason for this is the inherent complexity of DRO models and their comparatively recent development. Notably, a study by \cite{ghahtaraniworst} focused on moment-based ambiguity sets for the ALM problem and introduced worst-case Conditional Value at Risk (CVaR) as a risk measure to deal with parameter uncertainty in the problem. DRO has the potential to address some of the limitations of other methods, including the optimizer's curse in SP and the over-conservatism in RO. Moreover, DRO provides a way to explicitly consider the ambiguity in the distribution of financial variables.

In this paper, we aim to fill this gap in the literature by proposing DRO formulations for the ALM problem. We explore scenarios-based approaches to address the uncertainty of parameters in the ALM problem. 
Numerous studies have suggested that scenario-based analysis is superior to prediction-based analysis in financial problems. \cite{boender1997hybrid} argues that scenarios explicitly record assumptions about the future and provide a common framework for discussion. By utilizing scenarios, we can create a better understanding between managers and stakeholders, which can ultimately contribute to more effective decision-making. The main goal of this paper is to develop DRO scenario-based formulations and compare them against each other. The first formulation uses mixture ambiguity sets, each representing a convex combination of multiple distributions, each having multiple scenarios, which is commonly used in portfolio selection problems \citep{zhu2009worst}. 
In the second formulation, we investigate the case 
that the probabilities of scenarios are interval-bounded, but also there is a requirement that they add up to 1, which basically means that we are using a polyhedral set to represent the uncertain probabilities. 
Lastly, we incorporate the Wasserstein ambiguity set into the ALM problem, which is a metric-based ambiguity set. 
These ambiguity sets have been specifically chosen due to their suitability for utilization in scenario-based DRO formulation. 

We develop DRO models for an ALM problem that accounts for the ambiguity about the distribution of asset returns and interest rates. We also compare the performance of our DRO models to the traditional SP formulation of the ALM problem to demonstrate the advantages and limitations of each approach. By doing so, we hope to contribute to the development of a more robust and flexible framework for ALM that can better account for the uncertainty and variability in financial markets. 

The set of most used notations is shown in Table \ref{table:1paper2}, whereas the notations that are used once are defined in the text.

\begin{center}
\begin{longtable}{|p{0.2\textwidth} | p{0.7\textwidth}|} 
\caption{Notations and symbols}
\label{table:1paper2}
 \\ \hline 
 Symbol/Notation & Definition\\ [0.5ex] 
  \hline\hline
  \endhead
$t \in \{0,\dots,T\}$	&	Indices of decision moments	\\
\hline
\(T\)	&	Investment horizon	\\
\hline
$s \in \{1,\dots,S\}$	&	Indices of discount rate scenarios	\\
\hline
$k \in \{1,...,K\}$	&	Indices of asset return scenarios	\\
\hline
$y_t$ & Contribution rate at $t$, the fraction of sponsor and/or active employees' wages \\
\hline
\(y_{t,s}\)	&	Contribution rate at $t$ based on scenario $s$	\\
\hline
\(n \in \{0,\dots,N\}\)	&	 Indices of assets, where $n=0$ represents risk-free asset or cash  \\
\hline
\(x_{n,t}\)	&	Money invested in asset \(n\) at \(t\)	\\
\hline
\(x_{n,t,k}\)	&	Money invested in asset \(n\) at \(t\) based on scenario $k$	\\
\hline
\(A_{t}\)	&	Value of assets owned by the fund at \(t\)	\\
\hline
\(W_{t}\)	&	Wages earned by active members at \(t\)	\\
\hline
\(l_{t}\)	&	Payments made by the fund to retirees at \(t\)	\\
\hline
\(L_{t}\)	&	Net present value of liabilities of the fund at \(t\)	\\
\hline
\(L_{t,s}\)	& Net present value of liabilities of the fund at \(t\) based on scenario $s$	\\
\hline
\(\xi_{n,t}\)	&	Return on investment in asset \(n\) at \(t\)	\\
\hline
\(\xi_{n,t,k}\)	&	Return on investment in asset \(n\) at \(t\) based on scenario $k$	\\
\hline
\(\psi\)  &	  Minimum threshold of funding ratio	\\
\hline
\(\gamma\)  & Discount rate for calculating the present value	\\
\hline
\(W_{t}\)  & Net present value of wages at $t$	\\
\hline
\(p\)  & Discrete distribution function of discount rate	\\
\hline
\(q\)  & Discrete distribution function of asset returns	\\
\hline
\(P\)  & Ambiguity set of the distribution function of discount rate	\\
\hline
\(Q\)  & Ambiguity set of the distribution function of asset returns
\\ [1ex]
 \hline
\end{longtable}
\end{center}

The remaining sections of this paper are structured as follows. Section \ref{ALMproblem} introduces the mathematical formulation of the ALM problem. 
In Section \ref{secDROALM}, we present DRO formulations of the ALM problem based on the mixture distribution, the box, and the Wasserstein ambiguity sets. To test the proposed formulation, numerical experiments using real data from the Canada Pension Plan (CPP) are conducted, and the results are presented in Section \ref{resultspaper2}. Finally, Section \ref{Conclusionspaper2} offers some conclusions and suggests potential areas for future research. 

\section{ALM Model for Pension Funds}\label{ALMproblem}

The ALM problem under consideration aims to find an optimal investment strategy that achieves a trade-off between augmenting investment returns and reducing the risk of insolvency.
The objective of the ALM for a pension fund is to minimize the contribution rate by both the sponsor and active employees of the fund (\ie the contributors), as defined in previous studies \citep{bogentoft2001asset}. 
The financial burden is reduced by reducing the contribution rate, while the efficient investment strategy balances risk and returns over the investment horizon. The optimization process involves selecting the optimal portfolio of asset classes, such as stocks, bonds, and alternative investments, and the corresponding contribution rate for each period of the investment horizon. By finding the optimal solution to the ALM problem, the pension fund can ensure that it meets its future obligations while minimizing the financial burden on its stakeholders.

The investment horizon for the ALM problem under consideration, denoted as $T$, encompasses a series of decision moments represented by $t=0,\dots,T$. Several variables and decision-making components are at play in the ALM problem for pension funds. 
The contribution rate at $t$ is denoted by $y_{t}$, which is the fraction of the contributor's wage $w_t$ collected. Additionally, the decision variables $x_{n,t}$ represent the amount of money invested in asset $n$ at $t$, while $\xi_{n,t}$ is the return of asset $n$ in $t^{th}$ moment. The value of assets held by the fund at $t$ is represented by $A_t$, while the liabilities at that moment, which are payments made by the fund to retirees, are denoted by $l_t$. The present value of liabilities at $t$ is given by $L_t$, which is calculated by $\sum_{t=0}^T \frac{l_t}{(1+\gamma)^t},\; \forall t=0,\dots,T$. It is worth noting that in our study, benefit payments and liabilities are fixed and predefined, which classifies this type of pension fund as a \emph{defined-benefit plan}. The discount rate, $\gamma$, for the calculation of the present value of liabilities is a random variable. The funding ratio, a crucial parameter in the ALM problem, ensures that the ratio of assets owned by the fund to the present value of liabilities at $t$ is maintained above a minimum threshold $\psi$. This means that the fund has sufficient resources to meet its future obligations. Model \eqref{ALMpaper2} shows the mathematical formulation of the ALM problem:
\begin{subequations}\label{ALMpaper2}
\begin{alignat}{3}
& \min_{y_{t}, x_{n,t}}\quad&&{h(y_{1},\dots,y_{T})},\label{ALMpaper21}\\
&\;\;\text{s.t.}\quad&&\sum_{n=0}^{N}{x_{n,t}}=A_{t}+w_{t}y_{t}-l_{t},\quad&&t=0,\dots,T-1,\label{ALMpaper22}\\
&&& A_{t}\geq \psi L_{t},\quad&&t=1,\dots,T,\label{ALMpaper23}\\
&&& A_{t}=\sum_{n=0}^{N}{x_{n,t-1}}(1+\xi_{n,t}),\quad&&t=1,\dots,T,\label{ALMpaper24}\\
&&& x_{n,t} \in \mathcal{X}, y_{t}\in \mathcal{Y},\quad&&t=0,\dots,T,n=0,\dots,N\label{ALMpaper25}.
\end{alignat}
\end{subequations}

The objective function of the ALM problem \eqref{ALMpaper2}, introduced by \cite{bogentoft2001asset}, is denoted by $h(y_1,\dots,y_T)$, which is a function defined in terms of the contribution rate and plays a crucial role in determining the optimal ALM strategy. In particular, the objective function \eqref{ALMpaper21} can be defined as the present value of all contributions, \ie $h(y_1,\dots,y_T) = \sum_{t=1}^T W_{t}y_t$, where $W_{t}=\frac{w_{t}}{(1+\gamma)^t}$. The balance constraint \eqref{ALMpaper22} ensures 
that the sum of all investments at $t$ is equal to the assets held by the fund plus the contributions gathered at $t$ minus liabilities in that period. The funding ratio constraint \eqref{ALMpaper23} guarantees that the ratio of assets owned by the fund to the present value of liabilities at $t$ is greater than equal to a minimum threshold $\psi$. Constraint \eqref{ALMpaper24} describes how to calculate the value of assets owned by the fund at $t$ as a function of their values at $(t-1)$ and the return rates, $\xi_{n,t}$.
Any regulatory or practical (\eg nonnegativity) restrictions on the investment strategy and the contribution rates are encapsulated in the sets $\mathcal{X}$ and $\mathcal{Y}$, respectively, and are enforced by constraint \eqref{ALMpaper25}. Moreover, $A_0$, $w_0$, $y_0$, $l_0$, and $x_{n,0}$ represent the initial values of variables in the current period ($t= 0$) of the fund capital, employees’ wages, contribution rate, liabilities, and asset holdings, respectively. These parameters signify the starting condition of the fund, assumed to be known to the decision-maker.


To simplify the formulation, we express the objective function of model \eqref{ALMpaper2} using the vectors $\mathrm{W}=[W_{1},\dots,W_{T}]^\intercal\in \mathbb{R}^{T}$ and $\mathrm{y}=[y_{1},\dots,y_{T}]^\intercal\in \mathbb{R}^{T}$, which represent the present value of the contributors' wages and the contribution rate decision variables, respectively. The objective function can then be written as $\mathrm{W}^{\intercal}\mathrm{y}$.
We also introduce the vector $\mathrm{r}_{t}=\mathrm{e}+\upxi_t$ for $t=1,\dots,T$, where $\mathrm{e}$ is an all-ones vector of size $N+1$ and $\upxi_t$ is an uncertain vector that captures the variation in the plan's funding status. Additionally, we define the investment decision variable as a vector $\mathrm{x}_{t}=[x_{0,t},\dots,x_{n,t}]^\intercal$ for each $t$.
Using these notations, we can transform the ALM problem \eqref{ALMpaper2vector} into a vector representation as follows:
\begin{subequations}\label{ALMpaper2vector}
\begin{alignat}{3}
& \min_{\mathrm{y},\mathrm{x}_{t}}\quad&&\mathrm{W}^{\intercal}\mathrm{y},\label{ALMpaper2vector1}\\
&\;\;\text{s.t.}\quad&&\mathrm{e}^{\intercal}\mathrm{x}_{t}=\mathrm{r}_{t}^{\intercal}\mathrm{x}_{t-1}+w_{t}y_{t}-l_{t},\quad&& t=1,\dots,T-1,\label{ALMpaper2vector2}\\
&&&\mathrm{r}_{t}^{\intercal}\mathrm{x}_{t-1}\geq \psi L_{t},\quad&&t=1,\dots,T,\label{ALMpaper2vector3}\\
&&&\mathrm{x}_{t} \in \mathcal{X}, \mathrm{y}\in \mathcal{Y} \quad&&t=1,\dots,T\label{ALMpaper2vector4}.
\end{alignat}
\end{subequations}

The deterministic model \eqref{ALMpaper2vector} operates under the assumption of complete knowledge of all parameter values, such as the assets' rates of return $\upxi_t$, and the discount rate $\gamma$, which dictates the present value of future liabilities $L_t$. However, this assumption is unrealistic as these variables are inherently random.
To ensure the robustness of the model, it is important to account for the uncertainty associated with these parameters. In the following section, we present a DRO reformulation of \eqref{ALMpaper2vector} to address this issue.

\section{Distributionally Robust ALM}\label{secDROALM}

Before providing the DRO formulation, we begin by presenting a scenario-based SP formulation of the ALM problem. As previously discussed, the present value of wages, $\mathrm{W}$, and the present value of future liabilities, $L_{t}$, are both influenced by the uncertain discount rate, while all other parameters affecting them are assumed deterministic. As a result, they are perfectly correlated and thus can both be represented using a single set of discrete scenarios $\{s\}_{s=1,\dots,S}$ having a given distribution function $p(.)$. Likewise, we use a finite set of scenarios $\{k\}_{k=1,\dots,K}$, having the discrete distribution function $q(.)$, to capture the uncertainty of asset returns. With that, model \eqref{SPALMpaper2} presents an SP formulation of the ALM based on these scenario sets and distribution functions. Note that the additional subscript ($s$ or $k$) for the uncertain parameters denotes the scenario. 
\begin{subequations}\label{SPALMpaper2}
\begin{alignat}{3}
& \min_{\mathrm{y}, \mathrm{x}_{t}}\quad&&\mathbb{E}_{p}(\mathrm{W}_{s}^{\intercal}\mathrm{y}),\label{SPALMpaper21}\\
&\;\;\text{s.t.}\quad&&\mathrm{e}^{\intercal}\mathrm{x}_{t}=\mathbb{E}_{q}(\mathrm{r}_{t,k}^{\intercal}\mathrm{x}_{t-1})+w_{t}y_{t}-l_{t},\quad&& t=1,\dots,T-1,\label{SPALMpaper22}\\
&&&\mathbb{E}_{q}(\mathrm{r}_{t,k}^{\intercal}\mathrm{x}_{t-1})\geq \psi \mathbb{E}_{p}(L_{t,s}),\quad&&t=1,\dots,T,\label{SPALMpaper23}\\
&&&\mathrm{x}_{t} \in \mathcal{X}, \mathrm{y}\in \mathcal{Y} \quad&&t=1,\dots,T\label{SPALMpaper24}.
\end{alignat}
\end{subequations}

In model \eqref{SPALMpaper2}, it is essential to have comprehensive knowledge of the distribution functions for accurate analysis and decision-making. This requirement arises due to the significant impact of uncertainty in the problem domain. 
Additionally, the tractability of this problem is closely linked to the number of scenarios considered. As the number of scenarios increases, the complexity of the problem grows exponentially. Consequently, the computational burden and resource requirements for solving the problem also increase. Therefore, carefully considering the number of scenarios is crucial to balance accuracy and computational feasibility in tackling the model, which is a major limitation of this SP formulation.

However, in reality, these distribution functions may not be fully known, and therefore, we propose DRO as an alternative to SP for the ALM problem since the former does not require exact knowledge of the probability distributions. Moreover, DRO provides a robust solution by accounting for a range of possible different distributions, which enables decision-makers to hedge against various plausible distributional scenarios, leading to more reliable and stable solutions that are less sensitive to uncertain input parameters.
Consequently, we assume that the distribution functions belong to the sets that represent a range of possible probability distributions, called ambiguity sets. 
Let $P$ and $Q$ be ambiguity sets for the distribution functions of asset return and discount rate, respectively. Then, the DRO formulation of the ALM is presented in model \eqref{DROALMpaper2}:
\begin{subequations}\label{DROALMpaper2}
\begin{alignat}{3}
& \min_{\mathrm{y}, \mathrm{x}_{t}}\sup_{p \in P}\quad&&\mathbb{E}_{p}(\mathrm{W}_{s}^{\intercal}\mathrm{y}),\label{DROALMpaper21}\\
&\;\;\text{s.t.}\quad&&\mathrm{e}^{\intercal}\mathrm{x}_{t}=\inf_{q\in Q}\mathbb{E}_{q}(\mathrm{r}_{t,k}^{\intercal}\mathrm{x}_{t-1})+w_{t}y_{t}-l_{t},\quad&& t=1,\dots,T-1,\label{DROALMpaper22}\\
&&&\inf_{q\in Q}\mathbb{E}_{q}(\mathrm{r}_{t,k}^{\intercal}\mathrm{x}_{t-1})\geq \psi \sup_{p\in P}\mathbb{E}_{p}(L_{t,s}),\quad&&t=1,\dots,T,\label{DROALMpaper23}\\
&&&\mathrm{x}_{t} \in \mathcal{X}, \mathrm{y}\in \mathcal{Y} \quad&&t=1,\dots,T\label{DROALMpaper24}.
\end{alignat}
\end{subequations}

Model \eqref{DROALMpaper2} is based on the worst-case expected value of random variables.
For the remainder of the paper, we will explore various ambiguity sets that can be applied to the formulation presented in the model \eqref{DROALMpaper2}. 

\subsection{Mixture Distribution}\label{Mixture Distribution}
We are dealing with ambiguous discrete distribution functions, Whereas the scenarios themselves are deterministically defined. One approach to address this ambiguity of the distribution function is to use a set to represent the possible discrete distribution function. It is common to consider uncertain discrete distributions in portfolio selection problems (for more details, see \cite{costa2002robust}, \cite{ghaoui2003worst}, \cite{ghahtarani2022robust}). 
In this case, the ambiguity set is considered a mixture of predetermined likelihood distributions. Based on \cite{zhu2009worst},
these ambiguity sets are defined as $P_{M}:=\left\{p:p=\sum_{i=1}^{I}\lambda_{i}p^{i};\sum_{i=1}^{I}\lambda_{i}=1;\lambda_{i}\geq 0;i=1,\dots,I\right\}$, where $p^{i}$ is the $i^{th}$ likelihood distribution and $I$ is the number of likelihood distributions. 

Likewise, $Q_{M}:=\left\{q: q=\sum_{j=1}^{J}\lambda_{j}q^{j};\sum_{j=1}^{J}\lambda_{j}=1;\lambda_{j}\geq 0,;j=1,\dots,J\right\}$, where $q^{j}$ is the $j^{th}$ likelihood distribution and $J$ is the number of likelihood distributions. 
Although the ambiguity set $P_M$ includes all convex combinations of the $I$ likelihood distributions $p^i,\; i=1,\dots, I$, it is easy to show that the worst-case distribution for any given values of the decision variables is one of $I$ likelihood distributions themselves. This \emph{maximal solution} property is due to the fact that finding the worst-case distribution is analogous to solving a binary knapsack problem with unit-sized items and a knapsack capacity of 1, where only one item (having the highest value) is selected. A similar argument can be made for the ambiguity set $Q_M$, though with the lowest value item selected. Thus, to reformulate model \eqref{DROALMpaper2} with the mentioned ambiguity sets, we introduce the auxiliary variables $\theta$, $\mu_{t}$, and $\omega_{t}$, where \(\theta \geq \sum_{s=1}^{S}({\mathrm{W}_{s}^{\intercal}\mathrm{y})p_{s}^{i}}\), \(\mu_{t}\leq \sum_{k=1}^{K}(\mathrm{r}_{t,k}^{\intercal}\mathrm{x}_{t-1})q_{k}^{j}\), and \(\omega_{t}\geq \sum_{s=1}^{S}{L_{t,s}p_{s}^{i}}\). 
Using these ambiguity sets and auxiliary variables, the DRO problem can be written in the epigraph form as follows:
\begin{subequations}\label{MDROALMpaper2}
\begin{alignat}{3}
& \min_{\mathrm{y}, \mathrm{x}_{t}, \theta, \mu_{t}, \omega_{t}}\quad&&{\theta},\label{MDROALMpaper21}\\
&\;\;\text{s.t.}\quad&&\theta \geq \sum_{s=1}^{S}{(\mathrm{W}_{s}^{\intercal}\mathrm{y})p_{s}^{i}},\quad&& i=1,\dots,I,\label{MDROALMpaper22}\\
&&&\mathrm{e}^{\intercal}\mathrm{x}_{t}=\mu_{t}+w_{t}y_{t}-l_{t},\quad&& t=1,\dots,T-1,\label{MDROALMpaper23}\\
&&&\mu_{t} \geq \psi \omega_{t},\quad&&t=1,\dots,T,\label{MDROALMpaper24}\\
&&&\mu_{t}\leq \sum_{k=1}^{K}(\mathrm{r}_{t,k}^{\intercal}\mathrm{x}_{t-1})q_{k}^{j} \quad&&t=1,\dots,T,j=1,\dots,J,\label{MDROALMpaper25}\\
&&&\omega_{t}\geq \sum_{s=1}^{S}{L_{t,s}p_{s}^{i}},\quad&&t=1,\dots,T,i=1,\dots,I,\label{MDROALMpaper6}\\
&&&\mathrm{x}_{t} \in \mathcal{X}, \mathrm{y}\in \mathcal{Y} \quad&&t=1,\dots,T\label{MDROALMpaper7}.
\end{alignat}
\end{subequations}


Model \eqref{MDROALMpaper2} is a linear programming model, which is a tractable model and captures the ambiguity of discrete distribution functions. 

\subsection{Discrete Distribution with Box Ambiguity}\label{box ambiguity set}

The mixture-distribution ambiguity set proposed in subsection \ref{Mixture Distribution} has two main drawbacks. First, it confines the ambiguity about the discrete probability distributions to finite sets of elements (distribution functions) and their convex combinations while ignoring the possibility that the true distribution functions can take other forms. Although this issue can be partially alleviated by increasing the value of $I$, \ie using a large number of distribution functions that cover a wider range of possibilities, the problem size inevitably grows, thus reducing its tractability, which is the second drawback. Alternatively, a box ambiguity set can be used for the discrete distribution function, which provides does not need a large number of possible distribution functions in a convex set.
Note that for ambiguity sets of discrete distributions, model \eqref{DROALMpaper2} can be expanded to:
\begin{subequations}\label{BDROALMpaper2}
\begin{alignat}{3}
& \min_{\mathrm{y}, \mathrm{x}_{t}}\sup_{p \in P}\quad&&\sum_{s=1}^{S}(\mathrm{W}_{s}^{\intercal}\mathrm{y})p_{s},\label{BDROALMpaper21}\\
&\;\;\text{s.t.}\quad&&\mathrm{e}^{\intercal}\mathrm{x}_{t}=\inf_{q\in Q}\sum_{k=1}^{K}(\mathrm{r}_{t,k}^{\intercal}\mathrm{x}_{t-1})q_{k}+w_{t}y_{t}-l_{t},\quad&& t=1,\dots,T-1,\label{BDROALMpaper22}\\
&&&\inf_{q \in Q} \sum_{k=1}^{K} (\mathrm{r}_{t,k}^{\intercal}\mathrm{x}_{t-1})q_{k}\geq \psi \sup_{p \in P}\sum_{s=1}^{S}(L_{t,s})p_{s},\quad&& t=1,\dots,T,\label{BDROALMpaper23}\\
&&&\mathrm{x}_{t} \in \mathcal{X}, \mathrm{y}\in \mathcal{Y} \quad&&t=1,\dots,T\label{BDROALMpaper24}.
\end{alignat}
\end{subequations}


In model \eqref{BDROALMpaper2}, we are dealing with two ambiguity sets $P$ and $Q$ that contain probability distributions of the random variables. Specifically, $p(\cdot) \in P$, which is defined as $P:=\{p: p_s=p_{s}^{0}+\eta_{s}: \sum_{s=1}^{S}\eta_{s}=0, \underline{\eta}_{s}\leq \eta_{s}\leq \overline{\eta}_{s}\}$, where $p_{s}^{0}$ is the nominal probability of scenario $s$, and $\eta_{s}\in[\underline{\eta}_{s}, \overline{\eta}_{s}]$ is a bounded perturbation from it. Likewise, $q(\cdot) \in Q$, which is defined as $Q:=\{q(.):q_k=q_{k}^{0}+\xi_{k}: \sum_{k=1}^{K}\xi_{k}=0, \underline{\xi}_{k}\leq \xi_{k}\leq \overline{\xi}_{k}\}$, where $q_{k}^{0}$ is the nominal probability of scenario $k$, while $\xi_{k}\in [\underline{\xi}_{k}, \overline{\xi}_{k}]$. 

To reformulate model \eqref{BDROALMpaper2}, we apply LP duality to the inner problems. Specifically, the inner problem of the uncertain objective function \eqref{BDROALMpaper21} can be expressed as follows:
\begin{subequations}\label{objBDROALMpaper2}
\begin{alignat}{3}
& \max_{\eta_{s}}\quad&&{\sum_{s=1}^{S}(\mathrm{W}_{s}^{\intercal}\mathrm{y})(p^{0}_{s}+\eta_{s})},&&\quad\label{objBDROALMpaper21}\\
&\text{s.t.}\quad&&\sum_{s=1}^{S}\eta_{s}=0,\quad&&\quad(z),\label{objBDROALMpaper22}\\
&&&\eta_{s}\leq \overline{\eta}_{s},\quad&&\quad s=1,\dots,S,\;\;\;\;\;(d_{s}^{+}),\label{objBDROALMpaper23}\\
&&&-\eta_{s}\leq -\underline{\eta}_{s},\quad&&\quad s=1,\dots,S,\;\;\;\;\;(d_{s}^{-})\label{objBDROALMpaper24},
\end{alignat}
\end{subequations}
where $z$, $d_{s}^{+}$, and $d_{s}^{-}$ are the dual variables of their respective constraints. Problem \eqref{objBDROALMpaper2} aims to optimize over the perturbation parameters $\eta_{s}$. 
Thus, the dual form of model \eqref{objBDROALMpaper2} can be written as follows: 
\begin{subequations}\label{dualobjBDROALMpaper2}
\begin{alignat}{3}
&\sum_{s=1}^{S}(\mathrm{W}_{s}^{\intercal}\mathrm{y})p^{0}_{s}+\min_{d_{s}^{+}\geq 0, d_{s}^{-}\geq 0, z}\quad&&{\sum_{s=1}^{S}{(d_{s}^{+}\overline{\eta}_{s}-d_{s}^{-}\underline{\eta}_{s}})},\label{dualobjBDROALMpaper21}\\
&\;\;\text{s.t.}\quad&&z+d_{s}^{+}-d_{s}^{-}\geq \mathrm{W}_{s}^{\intercal}\mathrm{y},\quad&& s=1,\dots,S.\label{dualobjBDROALMpaper22}
\end{alignat}
\end{subequations}

Likewise, constraints \eqref{BDROALMpaper22} and \eqref{BDROALMpaper23} involve the inner optimization of uncertain parameters related to asset returns. Specifically, this inner optimization can be expressed as follows:
\begin{subequations}\label{conBDROALMpaper2}
\begin{alignat}{3}
& \min_{\xi_{k}}\quad&&{\sum_{k=1}^{K}(\mathrm{r}_{t,k}^{\intercal}\mathrm{x}_{t-1})(q^{0}_{k}+\xi_{k})},&&\quad\label{conBDROALMpaper21}\\
&\text{s.t.}\quad&&\sum_{k=1}^{K}\xi_{k}=0,\quad&&\quad(\Gamma),\label{conBDROALMpaper22}\\
&&&-\overline{\xi}_{k} \leq -\xi_{k},\quad&&\quad k=1,\dots,K,\;\;\;\;\;(\omega_{k}^{+}),\label{conBDROALMpaper23}\\
&&&\underline{\xi}_{k} \leq \xi_{k},\quad&&\quad k=1,\dots,K,\;\;\;\;\;(\omega_{k}^{-}),\label{conBDROALMpaper24}
\end{alignat}
\end{subequations}
where $\Gamma$, $\omega_{k}^{+}$, and $\omega_{k}^{-}$ are dual variables. The problem described in \eqref{conBDROALMpaper2} optimizes over the perturbation variables $\xi_{k}$. To achieve this goal, the objective function \eqref{conBDROALMpaper21} is reformulated as $\sum_{k=1}^{K}(\mathrm{r}_{t,k}^{\intercal}\mathrm{x}_{t-1})q^{0}+\min_{\xi_{k}}{\sum_{k=1}^{K}(\mathrm{r}_{t,k}^{\intercal}\mathrm{x}_{t-1})\xi_{k}}$, which expresses the minimum value of the linear combination $\sum_{k=1}^{K}(\mathrm{r}_{t,k}^{\intercal}\mathrm{x}_{t-1})\xi_{k}$ over all possible values of $\xi_{k}$. The dual form of the model presented in \eqref{conBDROALMpaper2} can be expressed as follows:
\begin{subequations}\label{dualconBDROALMpaper2}
\begin{alignat}{3}
&\sum_{k=1}^{K}(\mathrm{r}_{t,k}^{\intercal}\mathrm{x}_{t-1})q^{0}+ \max_{\omega_{k}^{+}\geq 0, \omega_{k}^{-}\geq 0, \Gamma}\quad&&{\sum_{k=1}^{K}{(\omega_{k}^{-}\underline{\xi}_{k}-\omega_{k}^{+}\overline{\xi}_{k}})},\label{dualconBDROALMpaper21}\\
&\;\;\text{s.t.}\quad&&\Gamma+\omega_{k}^{-}-\omega_{k}^{+}\leq \mathrm{r}_{t,k}^{\intercal}\mathrm{x}_{t-1},\quad&& k=1,\dots,K\label{dualconBDROALMpaper22}.
\end{alignat}
\end{subequations}

Finally, the inner optimization model related to uncertainty of \(L_{t,s}\) for each \(t\) in constraint \eqref{BDROALMpaper23} is as follows:
\begin{subequations}\label{libBDROALMpaper2}
\begin{alignat}{3}
& \max_{\eta_{s}}\quad&&{\sum_{s=1}^{S}{L_{ts}}(p^{0}_{s}+\eta_{s})},&&\quad\label{libBDROALMpaper21}\\
&\text{s.t.}\quad&&\sum_{s=1}^{S}\eta_{s}=0,\quad&&\quad(z),\label{libBDROALMpaper22}\\
&&& \eta_{s}\leq \overline{\eta}_{s},\quad&&\quad s=1,\dots,S,\;\;\;\;\;(d_{s}^{+}),\label{libBDROALMpaper23}\\
&&&-\eta_{s} \leq -\underline{\eta}_{s},\quad&&\quad s=1,\dots,S,\;\;\;\;\;(d_{s}^{-})\label{libBDROALMpaper24}.
\end{alignat}
\end{subequations}

The optimization problem \eqref{libBDROALMpaper2} is over $\eta_{s}$. Then the objective function \eqref{libBDROALMpaper21} is transformed to $\sum_{s=1}^{S}{L_{ts}}{p^{0}_{s}}+\max_{\eta_{s}}{\sum_{s=1}^{S}{L_{ts}}\eta_{s}}$.
The dual form of model \eqref{libBDROALMpaper2} for each \(t\) is as follows:
\begin{subequations}\label{duallibBDROALMpaper2}
\begin{alignat}{3}
&\sum_{s=1}^{S}{L_{ts}}{p^{0}_{s}}+\min_{d_{s}^{+}\geq 0, d_{s}^{-}\geq 0, z}\quad&&{\sum_{s=1}^{S}{(d_{s}^{+}\overline{\eta}_{s}-d_{s}^{-}\underline{\eta}_{s}})},\label{duallibBDROALMpaper21}\\
&\;\;\text{s.t.}\quad&&z+d_{s}^{+}-d_{s}^{-}\geq L_{t,s},\quad&& s=1,\dots,S\label{dualconBDROALMpaperp22}.
\end{alignat}
\end{subequations}

The DRO of the ALM model with box ambiguity set can be formulated by substituting the dual forms of the optimization problems \eqref{dualobjBDROALMpaper2}, \eqref{dualconBDROALMpaper2}, and \eqref{duallibBDROALMpaper2} into the original optimization problem expressed in \eqref{BDROALMpaper2}. The resulting final formulation is as follows: 

\begin{subequations}\label{finalBDROALMpaper2}
\begin{alignat}{3}
& \min_{\mathrm{y}, \mathrm{x}_{t}, d_{s}^{+}\geq 0, d_{s}^{-}, \omega_{k}^{+}, \omega_{k}^{-}\geq 0, \Gamma, z}\quad&&{\sum_{s=1}^{S}\left(\mathrm{W}_{s}^{\intercal}\mathrm{y}\right)p^{0}_{s}}+\sum_{s=1}^{S}{\left(d_{s}^{+}\overline{\eta}_{s}-d_{s}^{-}\underline{\eta}_{s}\right)},\label{finalBDROALMpaper21}
\end{alignat}
\begin{align}
&\;\;\text{s.t.}\quad&&\mathrm{e}^{\intercal}\mathrm{x}_{t}=\sum_{k=1}^{K}\left(\mathrm{r}_{t,k}^{\intercal}\mathrm{x}_{t-1}\right)q^{0}_{k}+\sum_{k=1}^{K}\left(\omega_{k}^{-}\underline{\xi}_{k}-\omega_{k}^{+}\overline{\xi}_{k}\right)+\nonumber\\
&&& w_{t}y_{t}-l_{t},\quad&& t=1,\dots,T-1,\label{finalBDROALMpaper22}
\end{align}
\begin{align}
&&& \sum_{k=1}^{K}\left(\mathrm{r}_{t,k}^{\intercal}\mathrm{x}_{t-1}\right)q_{k}^{0}+\sum_{k=1}^{K}{\left(\omega_{k}^{-}\underline{\xi}_{k}-\omega_{k}^{+}\overline{\xi}_{k}\right)}\geq\nonumber\\
&&& \psi\left(\sum_{s=1}^{S}{L_{t,s}p^{0}_{s}}+\sum_{s=1}^{S}{\left(d_{s}^{+}\overline{\eta}_{s}-d_{s}^{-}\underline{\eta}_{s}\right)}\right),\quad&& t=1,\dots,T,\label{finalBDROALMpaper23}
\end{align}
\begin{align}
&&&z+d_{s}^{+}-d_{s}^{-}\geq \mathrm{W}_{s}^{\intercal}\mathrm{y},\quad&& s=1,\dots,S,\label{finalBDROALMpaper24}\\
&&&\Gamma+\omega_{k}^{-}-\omega_{k}^{+}\leq \mathrm{r}_{t,k}^{\intercal}\mathrm{x}_{t-1} \quad&&t=1,\dots,T,k=1,\dots,K,\label{finalBDROALMpaper25}\\
&&&z+d_{s}^{+}-d_{s}^{-}\geq L_{t,s} \quad&&t=1,\dots,T,s=1,\dots,S.\label{finalBDROALMpaper26}  
\end{align}
\end{subequations}






Problem \eqref{finalBDROALMpaper2} is the tractable reformulation of \eqref{BDROALMpaper2} with box ambiguity sets. 

\subsection{Wasserstein Ambiguity Set}\label{Metric-based}

An ambiguity set that has drawn a lot of attention recently due to its favorable properties (\ie finite sample guarantee, asymptotic consistency, and tractability) is that based on the Wasserstein metric \citep{mohajerin2018data}. Unlike the mixture-distribution and box ambiguity sets utilized earlier, which consider only distributions that are supported on the same support set of the empirical distribution (\ie use the same set of scenarios), the Wasserstein ambiguity set includes all distributions, discrete or continuous, that are sufficiently close to the empirical distribution. 
Thus, it offers higher flexibility and a more realistic representation of the uncertainty of the random problem parameters. In other words, we do not only consider the ``original" scenarios on which the empirical distribution is supported, but also other scenarios not seen before. 


The Wasserstein ambiguity set can be constructed using the discrete empirical probability distribution $\hat{p}=\frac{1}{S}\sum_{s\in S}{\delta_{\hat{\mathrm{W}}_{s}}}$,
where $\delta$ is an indicator function that takes the value 1 for elements of the discrete set of scenarios $\hat{\Xi}_{\mathrm{W}}:={\hat{\mathrm{W}}_{1},...,\hat{\mathrm{W}}_{s}}\subset \Xi_{\mathrm{W}}$ and 0 elsewhere. 
Specifically, the ambiguity set is defined as $D(\hat{p},\epsilon^1)=\{p\in M \mid \begin{matrix}
P(\hat{\mathrm{W}}\in \Xi_{\mathrm{W}})=1 ,\
dw(\hat{p},p)\leq \epsilon^{1}
\end{matrix}\}$, where $dw(\hat{p},p)$ is the Wasserstein distance between the discrete empirical distribution $\hat{p}$ and a probability distribution $p$, and $\epsilon^1$ is the radius of the ambiguity set. This ambiguity set is designed to capture a range of probability distributions within a certain distance of the empirical distribution.
Similarly, the ambiguity set for the vector of asset returns random variables $\mathbf{r}_{t}$, denoted as $D(\hat{q}_{t},\epsilon_{t}^{2})$, is constructed using the discrete empirical probability distribution $\hat{q}_{t}=\frac{1}{K}\sum_{k\in K}{\delta_{\hat{\mathrm{r}}_{t}}}$. Here, $\hat{\Xi}_{\mathrm{r}_{t}}:={\hat{\mathrm{r}}_{t,1},...,\hat{\mathrm{r}}_{t,k}}\subset \Xi_{\mathrm{r}_{t}};\forall t$ is the set of empirical realizations of the vector of random variables $\mathbf{r}_{t}$. The ambiguity set $D(\hat{q}_{t},\epsilon_{t}^{2})$ is defined as $D(\hat{q}_{t},\epsilon_t^2)=\{q_{t} \in M \mid \begin{matrix}
P(\hat{\mathrm{r}_{t}}\in \Xi{\mathrm{r}_{t}})=1,\; \
dw(\hat{q}_{t},q_{t})\leq \epsilon_t^2
\end{matrix}\}$, where $dw(\hat{q}_{t},q_{t})$ is the Wasserstein distance between the discrete empirical distribution $\hat{q}_{t}$ and the probability distribution $q_{t}$, and $\epsilon_t^2$ is the radius of the ambiguity set.
To calculate the Wasserstein distance between two probability metrics $q_1$ and $q_2$, we use the integral representation $dw(q_1,q_2)=\int_{\Xi^{2}} ||\xi_{1},\xi_{2}||Q(d\xi_{1},d\xi_{2})$, where $Q$ is the joint distribution of $\xi_{1}$ and $\xi_{2}$ with marginal probabilities $q_{1}$ and $q_{2}$, respectively. This distance measure is used to capture the similarity between two probability distributions, where a smaller Wasserstein distance corresponds to a higher similarity between the distributions.
Based on \cite{mohajerin2018data}, under a convexity condition of the support set $\Xi_{\mathrm{W}}:=\{\mathbb{C} \mathrm{W}\leq \mathrm{d}\}$, where $\mathbb{C}\in \mathbb{R}^{m \times (t+1)}$, $\mathrm{d}\in \mathbb{R}^{m}$, constraint \eqref{DROALMpaper21} is transformed into:
\begin{subequations}\label{WDRBALMp1}
\begin{alignat}{3}
& \inf_{\lambda, \nu_{s}, \upgamma_{s}\geq 0}\quad&&{\lambda \epsilon^{1}+\frac{1}{S}\sum_{s\in S}{\nu_{s}}},\label{WDRBALMp11}\\
&\;\;\text{s.t.}\quad&&\hat{\mathrm{W}}^{\intercal}_{s}\mathrm{y}+(\mathrm{d}-\mathbb{C}\hat{\mathrm{W}}_{s})^{\intercal}\upgamma_{s}\leq \nu_{s},\quad&& s=1,\dots,S,\label{WDRBALMp12}\\
&&&\|\mathbb{C}^{\intercal}\upgamma_{s}-\mathrm{y}\|_{*}\leq \lambda,\quad&& s=1,\dots,S\label{WDRBALMp13}, 
\end{alignat}
\end{subequations}
where $\upgamma_{s} \in \mathbb{R}^{m}$, and $\|.\|_{*}$ is the dual norm of $\|.\|$, the norm used in the Wasserstein metric definition. With that, the right-hand side of constraint \eqref{DROALMpaper23}, which has the $p$-distributed random parameter $L_t$ with the support set $\Xi_{L_{t}}:=\{\mathrm{f}_{t} L_{t}\leq \mathrm{b}_{t},\;\forall t\in T\}$, where $\mathrm{f}_{t} \in \mathbb{R}^{m}$, $\mathrm{b}_{t} \in \mathbb{R}^{m}$ and $L_{t}\in \mathbb{R}$, reduces to:
\begin{subequations}\label{WDRBALM1}
\begin{alignat}{3}
& \inf_{\theta_{t}, \upsilon_{s,t}, \updelta_{s,t}\geq 0}\quad&&{\theta_{t} \epsilon_{t}^{1}+\frac{1}{S}\sum_{s\in S}{\upsilon_{s,t}}},\label{WDRBALM11}\\
&\;\;\text{s.t.}\quad&&\hat{L}_{s,t}+(\mathrm{b}_{t}-\mathrm{f}_{t}\hat{L}_{s,t})^{\intercal}\updelta_{s,t}\leq \upsilon_{s,t},\quad&& s=1,\dots,S,t=1,\dots,T,\label{WDRBALM12}\\
&&&\|\mathrm{f}_{t}^{\intercal}\updelta_{s,t}-1\|_{*}\leq \theta_{t},\quad&& s=1,\dots,S,t=1,\dots,T\label{WDRBALM3},
\end{alignat}
\end{subequations}
where $\updelta_{s,t} \in \mathbb{R}^{m}$. 

In constraints \eqref{DROALMpaper22} and \eqref{DROALMpaper22}, $\inf_{q\in Q}\mathbb{E}_{q}(\mathrm{r}_{t,k}^{\intercal}\mathrm{x}_{t-1})=-\sup_{q\in Q}\mathbb{E}_{q}(-\mathrm{r}_{t,k}^{\intercal}\mathrm{x}_{t-1})$. Then, by assuming that $\Xi_{\mathrm{r}_{t}}:=\{\mathbb{M}_{t} \mathrm{r}_{t} \leq \mathrm{u}_{t},\; t \in T\}$, where $\mathbb{M}_{t} \in \mathbb{R}^{m \times (n+1)}$ and $\mathrm{u}_{t} \in \mathbb{R}^{m}$, 
constraints \eqref{DROALMpaper22} and \eqref{DROALMpaper22} are reformulated as:
\begin{subequations}\label{WDRBALM2}
\begin{alignat}{3}
& \inf_{\phi_{t}, \varphi_{k,t}, \upzeta_{k,t}\geq 0}\quad&&{\phi_{t} \epsilon_{t}^{2}+\frac{1}{K}\sum_{k\in K}{\varphi_{k,t}}},\label{WDRBALM21}\\
&\;\;\text{s.t.}\quad&&-\hat{\mathrm{r}}_{k,t}^{\intercal}\mathrm{x}_{t-1}+(\mathrm{u}_{t}-\mathbb{M}_{t}\hat{\mathrm{r}}_{k,t})^{\intercal}\upzeta_{k,t}\leq \varphi_{k,t},\quad&& k=1,\dots,K,t=1,\dots,T,\label{WDRBALM22}\\
&&&\|\mathbb{M}^{\intercal}_{t}\upzeta_{k,t}-\mathrm{x}_{t-1}\|_{*}\leq \phi_{t},\quad&& k=1,\dots,K,t=1,\dots,T\label{WDRBALM23},
\end{alignat}
\end{subequations}
where $\upzeta_{k,t} \in \mathbb{R}^{m}$. By substituting \eqref{WDRBALMp1}, \eqref{WDRBALM1}, and \eqref{WDRBALM2} in model \eqref{DROALMpaper2}, we have the DRO counterpart of the ALM as follows:
\begin{subequations}\label{WDRBALMF}
\begin{alignat}{3}
&\;\; \inf_{\lambda, \nu_{s}, \upgamma_{s},\phi_{t}, \varphi_{k,t}, \upzeta_{k,t}, \theta_{t}, \upsilon_{s,t}, \updelta_{s,t}\geq 0}{\lambda \epsilon^{1}+\frac{1}{S}\sum_{s\in S}{\nu_{s}}},\label{WDRBALMF1}\\
&\;\;\text{s.t.}\;\;\;\; \mathrm{e}^{\intercal}\mathrm{x}_{t}=-\phi_{t} \epsilon_{t}^{2}-\frac{1}{K}\sum_{k\in K}{\varphi_{k,t}}+w_{t}y_{t,s}-l_{t},\;\;\;t=1,...,T-1,\label{WDRBALMF2}\\
&\;\;\;\;\;\;\;\;\;\; -\phi_{t} \epsilon_{t}^{2}-\frac{1}{K}\sum_{k\in K}{\varphi_{k,t}}\geq \theta_{t} \epsilon_{t}^{1}+\frac{1}{S}\sum_{s\in S}{\upsilon_{s,t}},\; t=1,\dots,T,\label{WDRBALMF3}\\
&\;\;\;\;\;\;\;\;\;\; \hat{\mathrm{W}}^{\intercal}_{s}\mathrm{y}+(\mathrm{d}-\mathbb{C}\hat{\mathrm{W}}_{s})^{\intercal}\upgamma_{s}\leq \nu_{s},\; s=1,\dots,S,\label{WDRBALMF4}\\
&\;\;\;\;\;\;\;\;\;\; \|\mathbb{C}^{\intercal}\upgamma_{s}-\mathrm{y}\|_{*}\leq \lambda,\; s=1,\dots,S,\label{WDRBALMF5}\\
&\;\;\;\;\;\;\;\;\;\; -\hat{\mathrm{r}}_{k,t}^{\intercal}\mathrm{x}_{t-1}+(\mathrm{u}_{t}-\mathbb{M}_{t}\hat{\mathrm{r}}_{k,t})^{\intercal}\upzeta_{k,t}\leq \varphi_{k,t},\; k=1,\dots,K, t=1,\dots,T,\label{WDRBALMF6}\\
&\;\;\;\;\;\;\;\;\;\; \|\mathbb{M}^{\intercal}_{t}\upzeta_{k,t}-\mathrm{x}_{t-1}\|_{*}\leq \phi_{t},\; k=1,\dots,K, t=1,\dots,T,\label{WDRBALMp23}\\
&\;\;\;\;\;\;\;\;\;\; \hat{L}_{s,t}+(\mathrm{b}_{t}-\mathrm{f}_{t}\hat{L}_{s,t})^{\intercal}\updelta_{s,t}\leq \upsilon_{s,t},\; s=1,\dots,S, t=0,\dots,T,\label{WDRBALMF7}\\
&\;\;\;\;\;\;\;\;\;\; \|\mathrm{f}_{t}^{\intercal}\updelta_{s,t}-1\|_{*}\leq \theta_{t},\; s=1,\dots,S, t=1,\dots,T,\label{WDRBALM13}\\
&\;\;\;\;\;\;\;\;\;\; \mathrm{x}_{t}\in \mathcal{X},\; \mathrm{y}\in \mathcal{Y}.\label{MDRBALM22}
\end{alignat}
\end{subequations}

The tractability of problem \eqref{WDRBALMF} depends on the dual norm $\|.\|_{*}$. We are using norm 2, $\|.\|_{2}$. Moreover, we use a box support set, where $\mathbb{M}=-\mathbb{I}$ and $\mathbb{C}=\mathbb{I}$, and $\mathcal{I}$ is the identity matrix that leads the box support sets for the uncertain parameters.

\section{An application to the Canadian Pension Plan}\label{resultspaper2}

For our study, we utilize data from the Canada Pension Plan (CPP) to conduct a series of numerical experiments. As a mandatory requirement for all employed Canadians aged 18-70, the CPP receives contributions from a vast majority of the working population. According to CPP's official website \footnote{https://open.canada.ca/data/en/dataset/1fab2afd-4f3c-4922-a07e-58d7bed9dcfc}, approximately 5.8 million individuals currently receive retirement benefits from CPP, with an average payout of $\$$811.21 in January 2023 \footnote{https://www.canada.ca/en/services/benefits/publicpensions/cpp/cpp-benefit/amount.html}. Additionally, CPP's investments report \footnote{https://www.cppinvestments.com/the-fund/our-performance/financial-results/f2022-annual-results} indicates that around 14,371,853 individuals are contributing to CPP.

CPP invests in a diverse portfolio of five asset classes, as per information from investing.com \footnote{https://ca.investing.com/}. These asset classes include fixed income, private equity, public equity, infrastructure, and real estate, which are geographically diversified in North America, Europe, and Asia. For our analysis, we have used data from ten major indexes from 2012 to 2022. The S\&P 500 index represents public equities, while the Private Equity Index (PRIVEXD) represents private equities. In addition, we use the SP/TSX Capped Real Estate Index (GSPRTRE) for the real estate sector, Treasury Yield 10 Years (TNX) for fixed-income assets, and S\&P Global Infrastructure TR (SPGTINTR) for infrastructure investment. The S\&P/TSX Composite is used as the index for the Canadian market, while the FTSEurofirst 300 represents public equities in Europe. For the private equity index in Europe, we use the STOXX Europe 20. The Shanghai Stock Exchange (SSE) and Nikkei-225 indexes have been utilized as representatives of investment in Asia.
As of 2022, the total value of assets under CPP management is estimated to be $\$$539 billion. Based on the most recent report from CPP \footnote{https://www.osfi-bsif.gc.ca/Eng/oca-bac/ar-ra/cpp-rpc/Pages/cpp30.aspx}, the projected earnings of contributors for 2022 have been calculated to be $\$$585,498 million, of which $\$$57,964 million (approximately 9.9\%) represents the contribution to CPP.

The inputs of our proposed models are the value of liabilities in each $t$, contributions of employees at each $t$, scenarios of asset returns, benefits paid by pension, and scenarios of interest rates. To generate scenarios of asset returns, we use Monte Carlo simulation based on the geometric Brownian motion (GBM), which is a common approach to generate random data in financial problems \citep{mcleish2011monte}. The formula for GBM is as follows:
\begin{align}\label{GBM}
\frac{\Delta Pr}{Pr}=\mu \Delta t+\sigma \epsilon \sqrt{\Delta t},  
\end{align}
where $Pr$ is the asset price, $\Delta Pr$ is the change in asset price, $\mu$ represents the expected return, $\sigma$ denotes the standard deviation of returns, $\epsilon$ is a normally-distributed random variable, and $\Delta t$ is the elapsed time period. 
Our analysis is based on the monthly returns of the ten mentioned indexes spanning from November 2012 to November 2022. During this period, we identified four distinct market regimes based on the long-term mean and standard deviation of the last 30 years. The first period, from November 2012 to February 2018, was characterized by steady growth with low volatility. The second period, from March 2018 to January 2020, experienced higher volatility than the previous period but still maintained a positive trend. The third period, from January 2020 to December 2021, was marked by high volatility and significant fluctuations. Finally, the post-pandemic period from January 2022 to November 2022 saw a return to high volatility, albeit with a different market trajectory. 
Using these historical data, we constructed different scenarios for our simulation. Our analysis, based on the $k$-mean clustering method of \cite{horvath2021clustering}, reveals that during the observed period, 51$\%$ of the time the market exhibited steady growth with low volatility (LV), 22$\%$ of the time it had medium volatility (MV) but still showed growth, 17$\%$ of the time it was characterized by increasing high volatility (IHV) with positive returns and 10$\%$ of the time it was a decreasing high volatility (DHV) decreasing market with negative returns. 
We generated 1000 scenarios for each asset in each period, corresponding to the 4 market regimes. The average return of these 1000 scenarios for each asset in each period, based on each market condition, is considered as the asset's return in that period for the 4 regimes.

The sets $\mathcal{X}$ and $\mathcal{Y}$ in model \eqref{DROALMpaper2} are defined using regulatory constraints based on the last decade's real investment structure in CPP. These constraints ensure that the contribution rate in each period falls within a range of $5\%$ to $10\%$, and the investment in the US market cannot exceed $60\%$ of the total fund, while the allocation to Canada must be at least $20\%$. Additionally, we mandate that a minimum of $10\%$ of the fund be invested in fixed-income assets. The allocation to Asia must not exceed $15\%$ of the fund, and the funding ratio must be at least $1.05$.

To evaluate the efficacy of our proposed DRO formulation, we conduct an out-of-sample performance analyses for four models. 
The first model is the mixture distribution ALM ($\mathtt{MD}$) \eqref{MDROALMpaper2} that incorporates four market conditions: LV, MV, IHV, and DHV. To determine the discrete probabilities associated with each market condition, we leverage the trends observed in historical data spanning the last 30 years. For the period from 2012 to 2022, the distribution functions of the market conditions are as follows: LV with a probability of 0.51, MV with a probability of 0.22, IHV with a probability of 0.17, and DHV with a probability of 0.1. Similarly, for the period from 2002 to 2012, the distribution functions are: LV with a probability of 0.35, MV with a probability of 0.39, IHV with a probability of 0.15, and DHV with a probability of 0.11. Lastly, for the period from 1992 to 2002, the distribution functions are: LV with a probability of 0.41, MV with a probability of 0.19, IHV with a probability of 0.20, and DHV with a probability of 0.06. Additionally, we consider a case where equal probabilities are assigned to each market condition, resulting in a probability of 0.25 for each LV, MV, IHV, and DHV. The second model is the box discrete distribution ALM ($\mathtt{BD}$) \eqref{finalBDROALMpaper2}, where half of the range of possible probabilities in each market condition is considered as volatility range in the box ambiguity set. The third mode is the Wasserstein metric ALM ($\mathtt{WM}$) \eqref{WDRBALMF}, where the radius of the Wasserstein ball is half of the range of possible probabilities in each market condition times to mean of asset return. The return on the market in the last 3 decades has been almost 10$\%$  \footnote{https://www.officialdata.org/us/stocks/s-p-500/2002}. Finally, the stochastic programming of ALM ($\mathtt{SP}$) is the last model for comparison to the proposed models.

\begin{figure}[h!]
    \begin{center}   \includegraphics[width=0.9\textwidth,height=0.50\textheight]{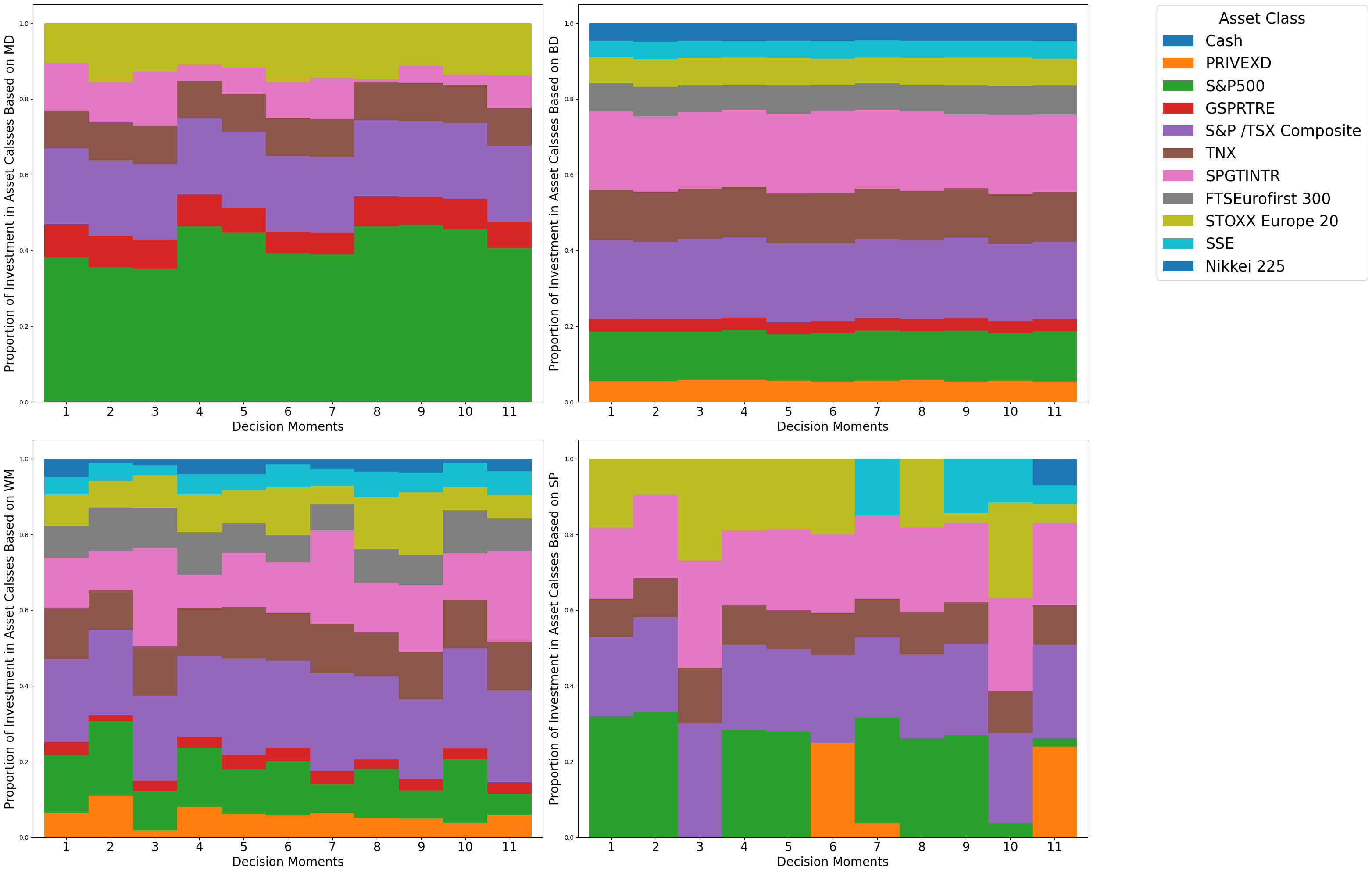}
     \caption{Comparision of optimal asset allocation}
     \label{fig:paper2inperf222}
     \end{center}
\end{figure}

Asset allocation is a crucial component in addressing the ALM problem. It requires determining how to distribute investments among different asset classes to achieve the desired return while minimizing risk. To compare the optimal asset allocation of four models over the investment horizon, Figure \ref{fig:paper2inperf222} has been presented. The horizontal axis represents the investment horizon, while the vertical axis depicts the proportion of investment in each asset class. The figure displays four models that correspond to the asset allocation of the $\mathtt{MD}$ model, the $\mathtt{BD}$ model, the $\mathtt{WM}$ model, and the $\mathtt{SP}$ model, respectively.


The Herfindahl-Hirschman Index (HHI) is a widely used metric to quantify market concentration. In portfolio selection, HHI involves assessing the diversification potential and risk exposure across different market segments or asset classes. A low HHI signifies a market characterized by greater diversification, where a larger number of assets distribute market share more evenly among themselves. In portfolio selection, lower HHIs can offer broader diversification benefits, potentially reducing overall portfolio risk by spreading exposure across a more extensive range of assets. In comparing the HHI across four models, $\mathtt{MD}$, $\mathtt{BD}$, $\mathtt{WM}$, and $\mathtt{SP}$, we observe notable distinctions. $\mathtt{MD}$ exhibits the highest average HHI at 0.256, suggesting a lower diversification. Following closely, $\mathtt{SP}$ presents an average HHI of 0.218, indicating a market structure with considerable concentration, albeit slightly lower than $\mathtt{MD}$. Conversely, $\mathtt{BD}$ and $\mathtt{WM}$ display lower average HHIs of 0.139 and 0.145 respectively, implying relatively more diversification with a greater number of asset classes sharing the portfolio more evenly. This comparison underscores varying degrees of diversification among the models. 

\begin{figure}[h!]
    \begin{center}   \includegraphics[width=0.9\textwidth,height=0.50\textheight]{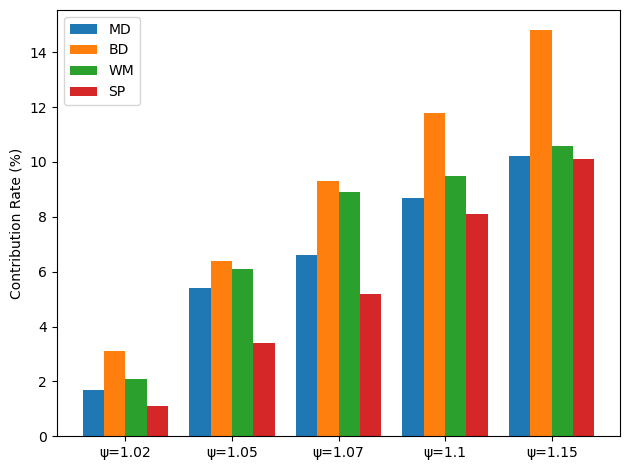}
     \caption{Optimal contribution rate for targeted funding ratios}
     \label{fig:paper2inperf2222}
     \end{center}
\end{figure}

The funding ratio ($\psi$) threshold is an important factor that affects the optimal contribution rate in ALM. Table \ref{tab:paper2my_label} and Figure \ref{fig:paper2inperf2222} present a comparison of the optimal contribution rates of four models ($\mathtt{MD}$, $\mathtt{BD}$, $\mathtt{WM}$, and $\mathtt{SP}$) under different $\psi$ values. 
Figure \ref{fig:paper2inperf2222} indicates that the optimal contribution rates for all models increase as the $\psi$ threshold increases. This is expected because a higher $\psi$ threshold implies a higher level of required funding, which in turn requires higher contribution rates to meet the threshold. Moreover, the table indicates that the optimal contribution rates for each model are different for different $\psi$ values. For instance, the $\mathtt{SP}$ model has the lowest optimal contribution rates among the four models for all $\psi$ values, while the $\mathtt{BD}$ model has the highest optimal contribution rates for $\psi$=1.15. Figure \ref{fig:paper2inperf2222} shows that the $\mathtt{WM}$ and $\mathtt{BD}$ models have relatively higher optimal contribution rates than the $\mathtt{MD}$ and $\mathtt{SP}$ models for all $\psi$ values. This suggests that the former models may be less conservative in managing mismatches between assets and liabilities and require higher contributions to ensure funding adequacy. 

\begin{table}[http]
\caption{Optimal contribution rates of four models based on funding ratio}
\centering
\begin{tabular}{|c||c|c|c|c|c|}
\hline
Models & $\psi$=1.02 & $\psi$=1.05 & $\psi$=1.07 & $\psi$=1.1 & $\psi$=1.15\\
\hline
\hline
$\mathtt{MD}$ & 1.7$\%$ & 5.4$\%$  & 6.6$\%$  & 8.7$\%$  & 10.2$\%$  \\
$\mathtt{BD}$ & 3.1$\%$  & 6.4$\%$  & 9.3$\%$  & 11.8$\%$  & 14.8$\%$  \\
$\mathtt{WM}$ & 2.1$\%$  & 6.1$\%$  & 8.9$\%$  & 9.5$\%$  & 10.6$\%$  \\
$\mathtt{SP}$ & 1.1$\%$  & 3.4$\%$  & 5.2$\%$  & 8.1$\%$  & 10.1$\%$  \\
\hline
\end{tabular}
\label{tab:paper2my_label}
\end{table}


We also evaluated the out-of-sample performance of the aforementioned models using a simulation to generate the testing data. 
Out-of-sample analysis refers to a method of evaluating the performance and robustness of a statistical or predictive model using data that is separate from the data used to develop or train the model. Specifically, we generated $1000$ scenarios of asset returns based on their distribution functions reported in \citep{ghahtaraniworst}. We then employed the optimal investment strategies of the $\mathtt{MD}$, $\mathtt{BD}$, $\mathtt{WM}$, and $\mathtt{SP}$ models to compare the average funding ratio and asset value in each period. The results of this analysis are presented in Table \ref{tab:paper2multicol112}. 

The out-of-sample analysis reveals that the DRO formulations employed in the ALM, $\mathtt{WM}$, and $\mathtt{BD}$ models exhibit superior performance compared to the $\mathtt{MD}$ and $\mathtt{SP}$ models.
Looking at the funding ratio for $\mathtt{MD}$, the ratio starts at 1.043 at the moment one, the current $t$, and decreases to 0.820 in period nine. However, the funding ratio then increases in $t$ 10 and 11, ending at 1.030. The fund return for $\mathtt{MD}$ starts at 0.001 at the current $t$, decreases to -0.002 at moment four, and then increases to 0.014 in period five. The fund return fluctuates between positive and negative values for the remaining periods.
For $\mathtt{BD}$, the funding ratio starts at 0.959 and gradually increases to 1.067 in period 11. The fund return for $\mathtt{BD}$ starts at 0.010 and steadily increases to 0.024 in period 11.
$\mathtt{WM}$ starts with a funding ratio of 0.972, which increases to 1.088 in period 10 and then decreases slightly to 1.114 in period 11. The fund return for $\mathtt{WM}$ increases from 0.014 to 0.025 in period 10 and then remains constant at 0.025 in period 11.
Finally, $\mathtt{SP}$ starts with a funding ratio of 0.981, which increases to 0.992 in period 11. The fund return for $\mathtt{SP}$ starts at 0.001, decreases to -0.007 in period three, and then increases to 0.004 in period 10. 

\begin{figure}[h!]
    \begin{center}   \includegraphics[width=0.9\textwidth,height=0.50\textheight]{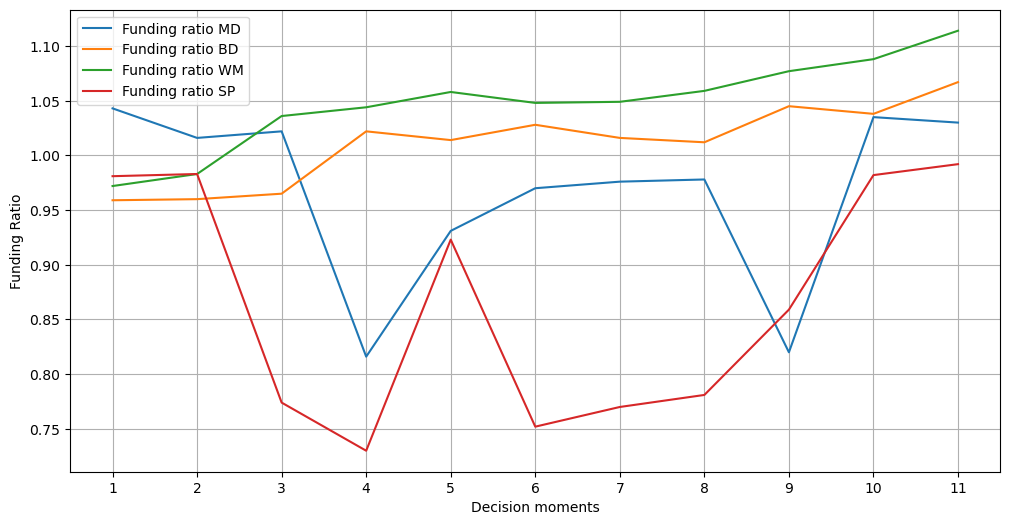}
     \caption{Funding Ratio of Different Models}
     \label{fig:paper2inperf22223}
     \end{center}
\end{figure}

\begin{figure}[h!]
    \begin{center}   \includegraphics[width=0.9\textwidth,height=0.50\textheight]{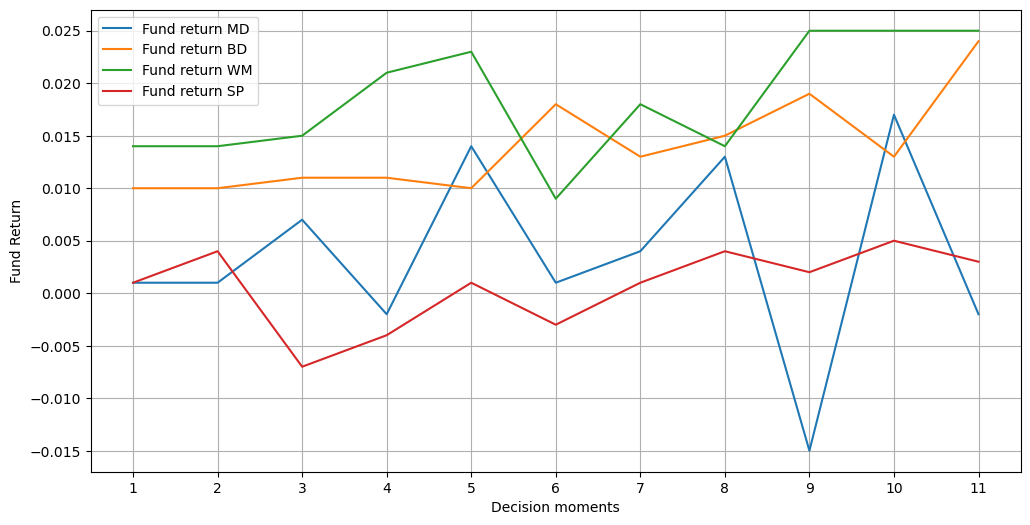}
     \caption{Fund Return of Different Models}
     \label{fig:paper2inperf22224}
     \end{center}
\end{figure}

The average funding ratio for $\mathtt{MD}$ is 0.976, indicating an average funding ratio below 1. This suggests that, on average, the assets are lower than the liabilities for $\mathtt{MD}$. $\mathtt{BD}$ has a mean funding ratio of 1.0115, indicating a slightly higher average ratio where assets are closer to liabilities. $\mathtt{WM}$ has the highest mean funding ratio among the models at 1.0480, suggesting a relatively higher average ratio of assets to liabilities. $\mathtt{SP}$ has the lowest mean funding ratio of 0.8661, indicating a lower average ratio where liabilities are higher than assets. In terms of the mean fund return, $\mathtt{MD}$ has a mean value of 0.0035, suggesting a slightly positive average return. $\mathtt{BD}$ has a higher mean fund return of 0.0140, indicating a relatively higher average return compared to $\mathtt{MD}$. $\mathtt{WM}$ has a mean fund return of 0.0185, suggesting a slightly higher average return among the models. $\mathtt{SP}$ has the lowest mean fund return value of 0.0006, indicating a very low average return.

The standard deviation values provide insights into the variability or dispersion of the performance metrics. For the funding ratio, $\mathtt{MD}$ has a standard deviation of 0.077, indicating a relatively higher variability compared to the other models. $\mathtt{BD}$ has the lowest standard deviation of 0.034, suggesting a lower variability in funding ratio performance. $\mathtt{WM}$ and $\mathtt{SP}$ have standard deviations of 0.039 and 0.102, respectively, indicating moderate to high variability. Regarding the fund return, $\mathtt{MD}$ has a standard deviation of 0.008, suggesting a relatively higher variability in returns. $\mathtt{BD}$ has the lowest standard deviation of 0.004, indicating a lower variability in fund return performance. $\mathtt{WM}$ has a standard deviation of 0.005, similar to $\mathtt{BD}$, suggesting relatively stable fund return outcomes. $\mathtt{SP}$ has a standard deviation of 0.003, indicating a lower variability compared to the other models.

\begin{table}
\caption{Pairwise comparison of means for funding ratio}
\begin{center}
\begin{adjustbox}{}
\small
\begin{tabular}{|c||c|c|c|}
    \hline
    Comparison & t-statistic & p-value \\
    \hline
    \hline
    $\mathtt{MD}$ vs $\mathtt{BD}$ & -1.6631 & 0.1119 \\
    $\mathtt{MD}$ vs $\mathtt{WM}$ & -2.9507 & 0.0079 \\
    $\mathtt{MD}$ vs $\mathtt{SP}$ & 2.4832 & 0.0220 \\
    $\mathtt{BD}$ vs $\mathtt{WM}$ & -2.2106 & 0.0389 \\
    $\mathtt{BD}$ vs $\mathtt{SP}$ & 4.2477 & 0.0004 \\
    $\mathtt{WM}$ vs $\mathtt{SP}$ & 5.2291 & 0.0000 \\
    \hline
\end{tabular}
\end{adjustbox}
\end{center}
\label{tab:funding_ratio_comparison}
\end{table}

The pairwise comparison of means for funding ratio, as presented in Table \ref{tab:funding_ratio_comparison}, reveals several notable distinctions between the ALM models. While no significant difference in means is observed between $\mathtt{MD}$ and $\mathtt{BD}$, $\mathtt{MD}$ exhibits a lower mean funding ratio compared to $\mathtt{WM}$, with a statistically significant difference confirmed by the negative t-statistic and p-value below 0.05. Conversely, $\mathtt{MD}$ demonstrates a higher mean funding ratio than $\mathtt{SP}$, again with statistical significance, indicated by the positive t-statistic and p-value below 0.05. $\mathtt{BD}$ displays a lower mean funding ratio compared to both $\mathtt{WM}$ and $\mathtt{SP}$, with statistically significant differences confirmed by the negative t-statistic and p-values below 0.05, respectively. Moreover, $\mathtt{BD}$ exhibits a higher mean funding ratio than $\mathtt{SP}$, with a notably high t-statistic and a highly significant p-value of 0.0004. $\mathtt{WM}$, on the other hand, exhibits a higher mean funding ratio than $\mathtt{SP}$, with an extremely significant difference confirmed by the positive t-statistic and p-value of 0.0000.

\begin{table}[htbp]
\caption{Pairwise comparison of means for fund return}
\begin{center}
\begin{adjustbox}{}
\small
\begin{tabular}{|c||c|c|c|}
    \hline
    \textbf{Comparison} & \textbf{t-statistic} & \textbf{p-value} \\
    \hline
    \hline
    $\mathtt{MD}$ vs $\mathtt{BD}$ & -3.4220 & 0.0027 \\
    $\mathtt{MD}$ vs $\mathtt{WM}$ & -4.6446 & 0.0002 \\
    $\mathtt{MD}$ vs $\mathtt{SP}$ & 0.9851 & 0.3363 \\
    $\mathtt{BD}$ vs $\mathtt{WM}$ & -2.0357 & 0.0552 \\
    $\mathtt{BD}$ vs $\mathtt{SP}$ & 7.4647 & 0.0000 \\
    $\mathtt{WM}$ vs $\mathtt{SP}$ & 8.7202 & 0.0000 \\
    \hline
\end{tabular}
\end{adjustbox}
\end{center}
\label{tab:fund_return_comparison}
\end{table}

The pairwise comparison of means for fund return, as detailed in Table \ref{tab:fund_return_comparison}, highlights significant differences between certain pairs of ALM models. Notably, $\mathtt{MD}$ exhibits statistically significant differences in mean fund return compared to both $\mathtt{BD}$ and $\mathtt{WM}$, as indicated by t-statistics of -3.4220 and -4.6446, respectively, and p-values below 0.05. Conversely, the comparison between $\mathtt{MD}$ and $\mathtt{SP}$ shows no statistically significant difference in mean fund return, with a t-statistic of 0.9851 and a p-value exceeding 0.05. $\mathtt{BD}$ demonstrates a significant difference in mean fund return compared to $\mathtt{SP}$, with a high t-statistic of 7.4647 and a very low p-value of 0.0000. Similarly, $\mathtt{WM}$ exhibits a significantly different mean fund return compared to $\mathtt{SP}$, with a higher t-statistic of 8.7202 and an equally low p-value of 0.0000.


\begin{table}
\caption{Out-sample performance of the ALM models}
\begin{center}
\begin{adjustbox}{width=1\textwidth}
\small
\begin{tabular}{|c||c|c||c|c||c|c||c|c|}
    \hline
    \multicolumn{1}{|c||}{}&\multicolumn{2}{c||}{$\mathtt{MD}$}&\multicolumn{2}{c||}{$\mathtt{BD}$}&\multicolumn{2}{c||}{$\mathtt{WM}$}&\multicolumn{2}{c|}{$\mathtt{SP}$}\\
    \cline{2-3}\cline{4-5}\cline{6-7}\cline{8-9}
    Decision moments &Funding ratio& Fund return &Funding ratio&Fund return&Funding ratio&Fund return&Funding ratio&Fund return\\  
    \hline
    \hline
\multirow{1}{*}{1}	&	1.043	&	0.001	&	0.959	&	0.010	&	0.972	&	0.014	&	0.981	&	0.001	\\
\multirow{1}{*}{2}	&	1.016	&	0.001	&	0.960	&	0.010	&	0.983	&	0.014	&	0.983	&	0.004	\\
\multirow{1}{*}{3}	&	1.022	&	0.007	&	0.965	&	0.011	&	1.036	&	0.015	&	0.774	&	-0.007	\\
\multirow{1}{*}{4}	&	0.816	&	-0.002	&	1.022	&	0.011	&	1.044	&	0.021	&	0.730	&	-0.004	\\
\multirow{1}{*}{5}	&	0.931	&	0.014	&	1.014	&	0.010	&	1.058	&	0.023	&	0.923	&	0.001	\\
\multirow{1}{*}{6}	&	0.970	&	0.001	&	1.028	&	0.018	&	1.048	&	0.009	&	0.752	&	-0.003	\\
\multirow{1}{*}{7}	&	0.976	&	0.004	&	1.016	&	0.013	&	1.049	&	0.018	&	0.770	&	0.001	\\
\multirow{1}{*}{8}	&	0.978	&	0.013	&	1.012	&	0.015	&	1.059	&	0.014	&	0.781	&	0.004	\\
\multirow{1}{*}{9}	&	0.820	&	-0.015	&	1.045	&	0.019	&	1.077	&	0.025	&	0.859	&	0.002	\\
\multirow{1}{*}{10}	&	1.035	&	0.017	&	1.038	&	0.013	&	1.088	&	0.025	&	0.982	&	0.005	\\
\multirow{1}{*}{11}	&	1.030	&	-0.002	&	1.067	&	0.024	&	1.114	&	0.025	&	0.992	&	0.003	\\
\multirow{1}{*}{Average}	&	0.976	&	0.0035	&	1.0115	&	0.0140	&	1.0480	&	0.0185	&	0.8661	&	0.0006	\\
\multirow{1}{*}{Std. Dev.}	&	0.077	&	0.008	&	0.034	&	0.004	&	0.039	&	0.005	&	0.102	&	0.003	\\

    \hline
\end{tabular}
\end{adjustbox}
\end{center}
\label{tab:paper2multicol112}
\end{table}

Based on the results presented in Table \ref{tab:paper2multicol112}, it is evident that the $\mathtt{WM}$ model outperforms the other models in terms of both fund return and funding ratio. With the highest mean values across the investment horizon, $\mathtt{WM}$ demonstrates superior performance from both perspectives while it has a moderate contribution rate in comparison to other models based on Table \ref{tab:paper2my_label}. In particular, the $\mathtt{WM}$ model exhibits a higher mean funding ratio compared to $\mathtt{MD}$ and $\mathtt{SP}$, indicating a more favorable financial position. This implies that $\mathtt{WM}$ manages to maintain a healthier balance between assets and liabilities, resulting in a greater ability to meet financial obligations. The higher funding ratio suggests a more robust and secure asset-liability management strategy.
Additionally, $\mathtt{WM}$ achieves the highest mean fund return among all the models, surpassing $\mathtt{BD}$, $\mathtt{MD}$, and $\mathtt{SP}$. This indicates that $\mathtt{WM}$ generates more favorable investment returns on average throughout the investment horizon. A higher mean fund return suggests better investment performance, potentially leading to higher profits and returns for the ALM strategy.
Therefore, based on the mean funding ratio and fund return values, it can be concluded that $\mathtt{WM}$ exhibits better performance than the other models in terms of stability and asset management. Its higher funding ratio indicates a stronger financial position, while the superior mean fund return reflects better investment outcomes. These findings highlight the effectiveness of the $\mathtt{WM}$ model in achieving both financial stability and favorable investment returns, making it a preferable choice among the options considered.

There is a technical reason why these models exhibit varying degrees of conservativeness and different out-of-sample performance. $\mathtt{SP}$ finds the optimal solution based on a single distribution function, specifically the empirical distribution function. As a result, it does not offer sufficient protection against unseen scenarios in the out-of-sample data. $\mathtt{MD}$ offers a solution by considering a broader range of possible distribution functions, as it takes into account the convex combination of multiple discrete distribution functions, forming a polyhedron. As a result, $\mathtt{MD}$ solutions are more conservative than $\mathtt{SP}$ solutions. On the other hand, $\mathtt{MD}$ solutions exhibit better out-of-sample performance compared to $\mathtt{SP}$ solutions. $\mathtt{BD}$ includes the widest range of distribution functions in its ambiguity set. Consequently, it offers the most conservative solution with the highest contribution rate, and highest opportunity cost. Finally, model $\mathtt{WM}$ uses a ball as its ambiguity set, covering a wide range of distribution functions. Although its contribution rate is not as high as that of 
$\mathtt{BD}$, it offers the most diversified portfolio and demonstrates better out-of-sample performance than the other models.

\section{Conclusions}\label{Conclusionspaper2}

Pension funds play a vital role in ensuring retirement income security for workers globally. However, they face challenges such as uncertainty of asset return and liability values. To address these issues, an effective asset-liability management (ALM) strategy must be implemented, balancing the competing objectives of generating returns and meeting future obligations. In this paper, we addressed the uncertainty of parameters in the ALM problem by exploring three different approaches: using mixture ambiguity sets with discrete scenarios and box uncertain discrete distribution functions. However, both of these approaches have limitations, and to overcome them, we incorporated the Wasserstein metric into the ALM problem. By incorporating the Wasserstein metric, we provide a more comprehensive and reliable approach to dealing with the limitations of ambiguity sets while maintaining the desirable properties of finite sample guarantee, asymptotic consistency, and tractability. 

This study has used data from the CPP to conduct a series of numerical experiments and tests to simulate different market scenarios and their impact on the plan. Monte Carlo simulation based on geometric Brownian motion was used to generate scenarios of asset returns. The analysis revealed four distinct market regimes during the observed period from November 2012 to November 2022.

Our analysis indicated that $\mathtt{BD}$ and $\mathtt{WM}$ offer more diversified portfolios. However, the contribution rate of $\mathtt{WM}$ is nearly identical to that of $\mathtt{SP}$ and $\mathtt{MD}$ across various values of $\psi$. In contrast, $\mathtt{BD}$ has the highest contribution rate, highlighting its conservative approach compared to the other models.



The out-of-sample analysis of the ALM models reveals that the $\mathtt{WM}$ model demonstrates superior performance compared to the $\mathtt{MD}$, $\mathtt{BD}$, and $\mathtt{SP}$ models in terms of both funding ratio and fund return.  
The superior performance of the $\mathtt{WM}$ model in terms of funding ratio and fund return can be attributed to its robust optimization approach and risk management capabilities. By incorporating worst-case distribution functions based on the scenarios of asset returns, the $\mathtt{WM}$ model is designed to handle extreme market conditions and mitigate potential risks.
The higher mean funding ratio of the $\mathtt{WM}$ model indicates a more conservative approach to asset-liability management, ensuring that the liabilities are well-covered by the available assets. This conservative stance provides a buffer against unexpected market fluctuations and reduces the likelihood of financial instability.

Furthermore, the higher mean fund return of the $\mathtt{WM}$ model suggests that it is able to capture profitable investment opportunities more effectively than the other models. This can be attributed to the optimization framework of the $\mathtt{WM}$ model, which aims to maximize investment returns while considering the constraints imposed by liabilities and risk tolerance.
The stability of the $\mathtt{WM}$ model's funding ratio and fund return is also evident from its lower standard deviation values compared to the other models. A lower standard deviation implies less variability and a more consistent performance over time. This stability is crucial for long-term financial planning and managing the risks associated with asset-liability mismatches.

Future research could focus on several directions to further enhance the understanding and application of ALM strategies for pension funds.
Firstly, it would be interesting to extend the analysis to consider the risk measures in the ALM problem. Incorporating such preferences into the ALM framework would allow for a more nuanced analysis of the trade-offs between risk and return and could provide insights into how different investors with varying risk preferences may adopt different ALM strategies.
Secondly, the current study used the Wasserstein metric to improve the robustness of the ALM models. However, there are other distance metrics that can be used to measure the discrepancy between probability distributions, such as the Kullback-Leibler divergence or the Total Variation distance. Future research could investigate the use of different distance metrics and compare their performance in the ALM context.
Lastly, the current study focused on the CPP, and it would be interesting to extend the analysis to other pension funds in different regions to explore the applicability and effectiveness of the ALM models in different contexts. Comparing the performance of ALM strategies across different pension funds could provide insights into how different market conditions and regulatory environments may impact the effectiveness of ALM strategies.





\bibliographystyle{apa-good}
\bibliography{main}

\end{document}